\documentstyle[aps,pra]{revtex}
\begin{document}
\draft
\title{cond-mat/9807046\\
Comparison of non-crossing perturbative approach and
generalized projection method for strongly coupled spin-fermion systems at
low doping.}
\author{R.O. Kuzian$^{\dagger }$}
\author{L.A. Maksimov$^{\ddagger }$}
\author{A.F.Barabanov, L.B. Litinskii$^{*}$}
\address{$^{\dagger }${\normalsize Institute for Materials Science,}\\
Krjijanovskogo 3, Kiev,\\
252180, Ukraine}
\address{$^{\ddagger }${\normalsize Russian Research Center Kurchatov Institute,}\\
Kurchatov sq.46, Moscow,\\
123182, Russia.}
\address{$^{*}${\normalsize Institute for High Pressure Physics,}\\
Troitsk, Moscow region,\\
142092, Russia.}
\date{}
\maketitle

\begin{abstract}
We analyze the two-dimensional spin-fermion model in the strong coupling
regime relevant to underdoped cuprates. We recall the set of general
sumrules that relate moments of spectral density and the imaginary part of
fermion self-energy with static correlation functions. We show that two-pole
approximation of projection method satisfies the sumrules for first four
moments of spectral density and gives an exact upper bound for quasiparticle
energy near the band bottom. We prove that non-crossing approximation that
is often made in perturbative consideration of the model violates the
sumrule for third moment of spectral density. This leads to wrong position
of lowest quasiparticle band. On the other hand, the projection method is
inadequate in weak coupling limit because of approximate treatment of
kinetic energy term. We propose a generalization of projection method that
overcomes this default and give the fermion self-energy that correctly
behaves both in weak and strong coupling limits.

Short title: Generalized projection method
\end{abstract}


\section{Introduction}

Over the past few years it became increasingly clear that anomalous normal
and superconducting state properties of high-$T_c$ cuprates are governed by
their close proximity to the transition into antiferromagnetic Mott-Hubbard
insulator state. The transition do occurs for strongly underdoped systems.
There are many indications that even overdoped cuprates are never far from
being antiferromagnetic \cite{Pines,Chub}. Such kinds of incipient
antiferromagnet at low temperatures have both fermionic and spin excitations
that strongly interact. For description of such interaction the
two-dimensional spin-fermion models are used. They have a general form
\begin{equation}
\label{s-f}\hat H=\hat H_{kin}+\hat H_s+\hat H_{int}.
\end{equation}
Here the kinetic energy term
$$
\hat H_{kin}=\sum_p\epsilon _pa_p^{\dagger }a_p,\quad a_p=N^{-1/2}\sum_ra_r
{\rm e}^{-ipr},
$$
describes the bare fermion propagation. We take for definiteness the
simplest version of nearest neighbor hopping $\epsilon _p=-2t(\cos p_x+\cos
p_y)$.

The spin subsystem may be described by microscopic Heisenberg Hamiltonian
$$
\hat H_s=\frac 12I\sum_{rg}S_{r+g}^\alpha S_r^\alpha ,
$$
or quite generally by a phenomenological form of dynamic susceptibility\cite
{MMP} $\chi (q,\omega )$, as far as we are interested by only fermion
spectrum. In the present work we suppose that spin subsystem is in the
paramagnetic state $\left\langle S_r^\alpha \right\rangle =0$ with strong
antiferromagnetic correlations $C_g=\left\langle S_{r+g}^\alpha S_r^\alpha
\right\rangle <0$ ,the correlation function $C_q=N^{-1}\sum_r\exp
(iqr)\left\langle S_{\rho +r}^\alpha S_\rho ^\alpha \right\rangle $ is
strongly peaked at wave vectors in the vicinity of $Q=(\pi ,\pi )$.

The third term in its simplest form describes Kondo interaction
\begin{equation}
\label{kondo}\hat H_{int}=J\sum_ra_r^{\dagger }\tilde S_ra_r=\frac J{\sqrt{N}
}\sum_{pq}a_p^{\dagger }\tilde S_qa_{p+q},\quad \tilde S_q=N^{-1/2}\sum_r
\tilde S_r{\rm e}^{iqr}.
\end{equation}

In the above formulae the sums run over the sites $r$ of a square lattice
and over the nearest neighbours with the lattice spacing $\mid g\mid =1$.
For short we miss the spin index for creation $a_{r\sigma }^{+}$ and
annihilation $a_{r\sigma }$ operators of the Fermi particles (we shall name
them as electrons) and in the Hamiltonian of Kondo-interaction $H_{int}$ we
take the notation $\tilde S_r=S_r^\alpha \sigma ^\alpha $; summation over
repeated indexes is understood everywhere; $\sigma ^\alpha $ are the Pauli
matrices. $\left\langle ...\right\rangle $ means thermodynamic average over
grand canonical ensemble.

In order to achieve the more realistic description, the interaction term may
be generalized by addition of nearest neighbour coupling $a_r^{\dagger }
\tilde S_ra_{r+g}+h.c.$ , etc. \cite{perlov}. It is essential that in any
case $\hat H_{int}$ remains local in real space and does not couple directly
the excitations on sites separated by large distance.

The usual approaches to the Hamiltonian (\ref{s-f}) exploit its apparent
similarity to that of polaron problem. Various perturbative methods have the
advantage of exact treatment of the one-particle part, $\hat H_{kin}$. It is
widely believed that non-crossing approximation \cite{KLR,perlov,Chub} is
appropriate for spin-fermion systems even in strong-coupling regime $J\gg t$
. Below we shall prove that in this regime the non-crossing approximation
violates the sum rule for the third moment of spectral density, as a result,
it gives the wrong position of lowest 'singlet' band. An alternative is the
Mori-Zwanzig projection technique \cite{mori}; due to the local nature of $
\hat H_{int}$ it suffices small number of basic operators for appropriate
account of local correlations. The obvious disadvantage of the technique is
that the kinetic energy is treated in approximate way, as a consequence, it
fails to correctly reproduce the weak-coupling regime $J\ll t$. In the
present paper we exactly take into account the kinetic energy term and use
the projection method for remaining terms. In the end we obtain the fermion
self-energy that correctly behaves in the weak-coupling limit and gives the
right lowest band position in strong coupling limit.

\section{Sum rules for Green's function and self-energy}

The quantities we calculate are the retarded fermion Green's function (GF)
\begin{equation}
\label{gfd}G_{XY}(\omega )=\left\langle X|Y^{\dagger }\right\rangle \equiv
-\imath \int_{t^{\prime }}^\infty \!\!dte^{\imath \omega (t-t^{\prime
})}\langle \{X(t),Y^{\dagger }(t^{\prime })\}\rangle \;.
\end{equation}
and the spectral density
$$
A_{XY}(\omega )=-\frac 1\pi {\rm Im}\left[ G_{XY}\left( \omega +\imath
0\right) \right] .
$$
Here and below $\{\ldots ,\ldots \},[\ldots ,\ldots ]$ stand for
anticommutator and commutator respectively. For diagonal GF $Y=X$, the
spectral density is positively definite $A_{XX}(\omega )>0$. Mori-Zwanzig
projection method allows to represent $G_{XX}(z),{\rm Im}z>0$ in the
continued fraction form \cite{forster,FuldeB}:
\begin{equation}
\label{cf}G_{XX}(z)=\frac{{\rm b}_0^2}{z-{\rm a}_0-}\frac{{\rm b}_1^2}{z-
{\rm a}_1-}\cdots \frac{{\rm b}_n^2}{z-{\rm a}_n-}\cdots
\end{equation}

where
\begin{equation}
\label{a0A}{\rm b}_0^2=\int_{-\infty }^{+\infty }A_{XX}(\omega )d\omega
,\quad {\rm a}_0=\frac 1{{\rm b}_0^2}\int_{-\infty }^{+\infty }\omega
A_{XX}(\omega )d\omega .
\end{equation}

The coefficients ${\rm b}_n,{\rm a}_n,n>0$ are related with the spectral
density $A_{XX}(\omega )$ via the set of orthogonal polynomials $P_n(\omega
) $, satisfying the recurrence \cite{rm3,rm5}:
$$
P_{-1}(\omega )=0,\quad P_0(\omega )=1,
$$

\begin{equation}
\label{polrec}P_{n+1}(\omega )=(\omega -{\rm a}_n)P_n(\omega )-{\rm b}
_n^2P_{n-1}(\omega ),
\end{equation}
and
\begin{equation}
\label{polb}{\rm b}_{n+1}^2=\frac{\int_{-\infty }^{+\infty }P_{n+1}^2(\omega
)A_{XX}(\omega )d\omega }{\int_{-\infty }^{+\infty }P_n^2(\omega
)A_{XX}(\omega )d\omega },
\end{equation}
\begin{equation}
\label{pola}{\rm a}_{n+1}=\frac{\int_{-\infty }^{+\infty }\omega
P_{n+1}^2(\omega )A_{XX}(\omega )d\omega }{\int_{-\infty }^{+\infty
}P_{n+1}^2(\omega )A_{XX}(\omega )d\omega }.
\end{equation}

Here we have used the nonnormalized form of the polynomials $\int_{-\infty
}^{+\infty }P_n(\omega )P_s(\omega )A_{XX}(\omega )d\omega =\delta
_{ns}(\prod_{m=1}^{m=n}{\rm b}_m)^2$.

On the other hand, from the equation of motion
\begin{equation}
\label{eqmot}\omega \left\langle X|Y^{\dagger }\right\rangle =\left\langle
\{X,Y^{\dagger }\}\right\rangle +\left\langle X{\cal L}|Y^{\dagger
}\right\rangle ,\quad X{\cal L}\equiv \left[ X,\hat H\right]
\end{equation}
follows the sum rule
\begin{equation}
\label{srule}\int_{-\infty }^{+\infty }F(\omega )A_{XY}(\omega )d\omega
=\left\langle \{XF({\cal L}),Y^{\dagger }\}\right\rangle
\end{equation}
for arbitrary function $F({\cal L})$. In particular, it establishes the
relations of coefficients ${\rm a}_n,{\rm b}_n$ with static correlation
functions
\begin{equation}
\label{b0a0}{\rm b}_0^2=\left\langle \{X,X^{\dagger }\}\right\rangle ,\quad
{\rm a}_0=\frac{\left\langle \{X{\cal L},X^{\dagger }\}\right\rangle }{
\left\langle \{X,X^{\dagger }\}\right\rangle },
\end{equation}
\begin{equation}
\label{b1a1}{\rm b}_1^2=\frac{\left\langle \{X({\cal L}-{\rm a}
_0)^2,X^{\dagger }\}\right\rangle }{\left\langle \{X,X^{\dagger
}\}\right\rangle },\quad {\rm a}_1=\frac{\left\langle \{X{\cal L}({\cal L}-
{\rm a}_0)^2,X^{\dagger }\}\right\rangle }{\left\langle \{X({\cal L}
-a_0)^2,X^{\dagger }\}\right\rangle }
\end{equation}

Now introducing the self energy $\Sigma (z)$ through the relation
\begin{equation}
\label{gfsigm}\left( z-\frac{\left\langle \{X{\cal L},X^{\dagger
}\}\right\rangle }{\left\langle \{X,X^{\dagger }\}\right\rangle }-\Sigma
(z)\right) G_{XX}(z)=\left\langle \{X,X^{\dagger }\}\right\rangle
\end{equation}
and comparing (\ref{cf}) and (\ref{gfsigm}) we see that $\Sigma (z)$ it is
the continued fraction similar to $G(z)$ . Thus we can introduce the
spectral density

$$
\rho (\omega )=-{\rm Im}[\Sigma (\omega +\imath 0^{+})]/\pi
$$
and obtain for it sum rules that follow from (\ref{srule})
\begin{equation}
\label{sigsrule}{\rm b}_1^2=\int_{-\infty }^\infty \rho (\omega )d\omega ,
\end{equation}
\begin{equation}
\label{sigsrule2}{\rm a}_1=\frac 1{{\rm b}_1^2}\int_{-\infty }^{+\infty
}\omega \rho (\omega )d\omega .
\end{equation}
As it follows from (\ref{b1a1}) and (\ref{pola}) the last equality relates
the third moment of $A_{XX}(\omega )$ , the first moment of $\rho (\omega )$
and static correlation functions. For spin-fermion models in the limit of
low doping, the spin-spin correlation functions are only involved in (\ref
{b1a1}). Below we show that non-crossing approximation violates sumrule (\ref
{sigsrule2}) and is obviously wrong in strong coupling limit $J\gg t$.

\section{Projection technique}

In practice, projection technique calculations are possible only for finite
basis set and only finite number of continued fraction levels is possible to
calculate in (\ref{cf}).

In the frameworks of our model we have
\begin{equation}
\label{emgf}\left( \omega -\epsilon _p\right) \left\langle a_p|a_p^{\dagger
}\right\rangle =1+J\sqrt{f_2}\left\langle b_p|a_p^{\dagger }\right\rangle
,\quad b_p=N^{-1/2}\sum_rb_r{\rm e}^{-ipr},
\end{equation}
$$
b_r=\frac 1{\sqrt{f_2}}\tilde S_ra_r,\quad f_2=\left\langle \tilde S_r\tilde
S_r\right\rangle =\frac 34.
$$
Thus the 'bare' electron operator $a_r$ with one-site spin polaron operator $
b_r$ represent the natural basis set for appropriate account of local
correlations. It is important that this set is closed relative to $\hat H
_{int}$ , i.e.
\begin{equation}
\label{emJ}\left[ a_r,\hat H_{int}\right] =J\sqrt{f_2}b_r,\quad \left[ b_r,
\hat H_{int}\right] =J\left( \sqrt{f_2}a_r-b_r\right) .
\end{equation}
Now the commutation relation
\begin{equation}
\label{emb}\left[ \tilde S_{r+R}a_r,\hat H\right] =-t\sum_g\tilde S
_{r+R}a_{r+g}+J\tilde S_{r+R}\tilde S_ra_r+\left[ \tilde S_{r+R}a_r,\hat H
_s\right]
\end{equation}
is projected onto the basis set in order to decouple the equation of motion
for higher order GF $G_b(p,\omega )\equiv \left\langle b_p|a_p^{\dagger
}\right\rangle .$ In the following, we neglect the spin excitation energy $
I\ll t,J$
\begin{equation}
\label{emgfb}\omega G_b=\left\langle \left[ b_p,\hat H\right] |a_p^{\dagger
}\right\rangle \simeq \left( \frac{C_g}{f_2}\epsilon _p-J\right) G_b+J\sqrt{
f_2}G_a.
\end{equation}
This gives both GF in two pole approximation
\begin{equation}
\label{twopole}G_{a,b}^{(2)}(p,\omega )=\frac{\left| \alpha _{a,b}^S\right|
^2}{\omega -\Omega _S}+\frac{\left| \alpha _{a,b}^T\right| ^2}{\omega
-\Omega _T},\ {\left| \alpha _{a,b}^S\right| ^2}+{\left| \alpha
_{a,b}^T\right| ^2}=1,
\end{equation}
here $\Omega _n,\alpha _i^n,i=a,b,n=S,T,\quad (\Omega _S<\Omega _T)$ are
eigenvalues and eigenvectors of the problem
\begin{equation}
\label{eipro}\left(
\begin{array}{cc}
{\rm a}_0-\Omega _n & {\rm b}_1 \\ {\rm b}_1 & {\rm a}_1-\Omega _n
\end{array}
\right) \left(
\begin{array}{c}
\alpha _1^n \\
\alpha _2^n
\end{array}
\right) =0,
\end{equation}
where matrix elements are
\begin{equation}
\label{matel}{\rm a}_0=\epsilon _p,\quad {\rm b}_1=J\sqrt{f_2},\quad {\rm a}
_1=\frac{C_g}{f_2}\epsilon _p-J.
\end{equation}
Together with the normalization ${\rm b}_0^2=1$ the matrix elements
correspond to continued fraction coefficients of $G_a^{(2)}$ that coincide
with the first two pairs of coefficients of exact GF $G_a$. It means that $
G_a^{(2)}$ automatically satisfies the sumrules (\ref{b0a0}), (\ref{sigsrule}
) and (\ref{sigsrule2}).

Near the band bottom $G_a$ and $A_{aa}$ should have the form
\begin{equation}
\label{QP}G_a(p,\omega )=\frac{Z_a(p)}{\omega -E_p}+G_{inc},
\end{equation}
$$
A_{aa}(p,\omega )=Z_a(p)\delta (\omega -E_p)+A_{inc}(p,\omega ).
$$
Here $E_p$ and $Z(p)<1$ are quasiparticle energy and pole strength
respectively. The incoherent part. $A_{inc}$ is not zero for $\omega >\omega
_{\min }>E_p$. Now it is easy to show that $\Omega _S$ represents an {\em
exact upper bound} for $E_p$. Let us consider eigen operator for lowest
'singlet' band
$$
\xi _S=\alpha _a^Sa_p+\alpha _b^Sb_p.
$$
The GF
$$
G_{\xi \xi }=\left\langle \xi _S|\xi _S^{\dagger }\right\rangle =\frac{Z_\xi
(p)}{\omega -E_p}+G_{\xi \xi ,inc}
$$
has the pole at the same energy as bare fermion GF $G_a$ (in our model $
\alpha _a^S\neq 0$ for all $p$). On the other hand, from (\ref{a0A}) and (
\ref{eipro}) we have
\begin{equation}
\label{polepos}\Omega _S=Z_\xi (p)E_p+\int_{\omega _{\min }}^\infty \omega
A_{\xi \xi ,inc}(p,\omega )d\omega =Z_\xi (p)E_p+\left[ 1-Z_\xi (p)\right]
\Omega _{inc},
\end{equation}
here
$$
\Omega _{inc}\equiv \frac{\int_{\omega _{\min }}^\infty \omega A_{\xi \xi
,inc}(p,\omega )d\omega }{\int_{\omega _{\min }}^\infty A_{\xi \xi
,inc}(p,\omega )d\omega }\geq \omega _{\min }
$$
is the center of gravity of the incoherent part. As the pole strength is $
0\leq Z_\xi (p)\leq 1$, we have
\begin{equation}
\label{bound}E_p\leq \Omega _S\leq \Omega _{inc},
\end{equation}
i.e. for any $p$ the exact energy $E_p$ lies {\em always lower} then the
energy given by two pole approximation.

In the strong coupling limit that gives
\begin{equation}
\label{strcup}\Omega _S\approx -\frac 32J+\epsilon _p(\frac 14+C_g),
\end{equation}
and from (\ref{polepos}) it follows that actual pole position is lower than $
\Omega _S$. It is not difficult to calculate next continued fraction
coefficient
\begin{equation}
\label{b2}{\rm b}_2^2=\frac 1{f_2N}\sum_q\left( \epsilon _{p+q}-\frac 43
C_g\epsilon _p\right) ^2C_q\approx \left( \epsilon _{p+Q}-\frac 43
C_g\epsilon _p\right) ^2
\end{equation}
in the approximate equality we took into account that main contribution to
the sum over $q$ comes from the vicinity of $Q$ and $N^{-1}\sum_qC_q=f_2$.
We see that ${\rm b}_2$ is of the order of kinetic energy $t\ll J$. It means
that small polaron formed by our basic operators interact with the spin
subsystem much weaker than the bare hole. So, the expected polaron energy
renormalization from $\Omega _S$ to $E_p$ is of the order of $t^2/J$.

\section{Exact treatment of the kinetic energy}

From the above consideration it follows that the two-pole expression (\ref
{twopole}) is useful in strong coupling limit. In the opposite case $J\ll t$
it becomes not appropriate because it gives the self-energy in the one pole
form
\begin{equation}
\label{sig2}\Sigma ^{(2)}(z)={\rm b}_1^2/(z-{\rm a}_1)
\end{equation}
and cannot describe the damping of quasiparticle. The reason is in the
approximate treatment of the kinetic energy term in course of the projection
of the equation for $G_b$ ( \ref{emgfb}). Here we propose a generalization
of projection technique that completely removes this default and makes it
possible to exactly take into account the kinetic energy term.

We express $b_p$ in the following form
\begin{equation}
\label{em1}b_p=\frac 1{\sqrt{Nf_2}}\sum_q\tilde S_qa_{p+q},
\end{equation}
the equation of motion for every item gives
\begin{equation}
\label{em2}\left( \omega -\epsilon _{p+q}\right) \left\langle \tilde S
_qa_{p+q}|a_{p_1}^{\dagger }\right\rangle =\left\langle \left[ \tilde S
_qa_{p+q},\hat V\right] |a_{p_1}^{\dagger }\right\rangle .
\end{equation}
Here and below we denote $\hat V=\hat H_{int}+\hat H_s$. Now we project the
higher-order operator in the right-hand side of (\ref{em2} ) onto our basis
operators $B_{1,p}\equiv a_p,$ $B_{2,p}\equiv b_p$

$$
\left\langle \left\{ \left[ \tilde S_qa_{p+q},\hat V\right]
,B_{i,p}^{\dagger }\right\} \right\rangle =\frac 1{N\sqrt{N}}
\sum_{r_1r_2r_3}\left\langle \left\{ \left[ \tilde S_{r_1}a_{r_2},\hat V
\right] ,B_{i,r_3}^{\dagger }\right\} \right\rangle \exp \left( \imath
qr_1-\imath (p+q)r_2+\imath pr_3\right) =
$$
$$
\frac 1{N\sqrt{N}}\sum_{r_1r_2r_3}\left\langle \left\{ \left[ \tilde S_{r_1},
\hat V\right] a_{r_2}+\tilde S_{r_1}\left[ a_{r_2},\hat V\right]
,B_{i,r_3}^{\dagger }\right\} \right\rangle \exp \left( \imath q\left(
r_1-r_2\right) +\imath p\left( r_3-r_2\right) \right) =
$$
\begin{equation}
\label{ksq}\frac 1{\sqrt{N}}\sum_R\left\langle \left\{ \left[ \tilde S_{r+R},
\hat V\right] a_r+\tilde S_{r+R}\left[ a_r,\hat V\right] ,B_{i,r}^{\dagger
}\right\} \right\rangle \exp \left( \imath qR\right) =\frac 1{\sqrt{N}}
\sum_RK_{i,R}\exp \left( \imath qR\right) \equiv \frac 1{\sqrt{N}}K_{i,q},
\end{equation}
$i=1,2$ . We have used the local character of operator $\hat V$ that gives $
\delta _{r_{2,}r_3}$. Thus Eq.(\ref{em2}) may be rewritten as
\begin{equation}
\label{decmax}\left( \omega -\epsilon _{p+q}\right) \left\langle \tilde S
_qa_{p+q}|a_{p_1}^{\dagger }\right\rangle \simeq \frac 1{\sqrt{N}}
\sum_iK_{i,q}\left\langle B_{i,p}|a_{p_1}^{\dagger }\right\rangle .
\end{equation}
Now equation for $b_p$ is
\begin{equation}
\label{bp}\left\langle b_p|a_{p_1}^{\dagger }\right\rangle =\frac 1{\sqrt{
Nf_2}}\sum_q\left\langle \tilde S_qa_{p+q}|a_{p_1}^{\dagger }\right\rangle =
\frac 1{N\sqrt{f_2}}\sum_{i,q}\frac{K_{i,q}}{\left( \omega -\epsilon
_{p+q}\right) }\left\langle B_{i,p}|a_{p_1}^{\dagger }\right\rangle .
\end{equation}
Explicit calculation gives ($I\approx 0$)
$$
K_{1,q}=JC_q,\quad K_{2,q}=-\frac J{\sqrt{f_2}}C_q
$$
and we obtain the following form for the fermion self-energy
\begin{equation}
\label{sigMax}\Sigma \left( p,\omega \right) =\frac{J^2f_2}{J+f_2D_p^{-1}}
,\quad D_p(\omega )\equiv \frac 1N\sum_q\frac{C_q}{\left( \omega -\epsilon
_{p+q}\right) }.
\end{equation}
In the weak coupling limit $J\ll t$ this expression coincides with the
second order result of perturbation theory
\begin{equation}
\label{sigpert}\Sigma _{pert}\left( p,\omega \right) =J^2D_p(\omega ),
\end{equation}
which corresponds to the projection (\ref{ksq}) only on the first operator $
a_p$. In this limit a polaron of large radius is formed and bare fermion
represents slightly damping quasiparticle.

Now let us show that the self-energy (\ref{sigMax}) also give correct result
in the strong coupling limit. The pole position is given by $E_p-\epsilon
_p-\Sigma \left( p,E_p\right) =0$ . For determination of lowest band
position we may neglect the $\epsilon _{p+q}\propto t$ compared $\omega
\propto -J$ and write $D_p(\omega )\approx f_2/\omega $ , then
$$
\Sigma \left( p,\omega \right) \approx \frac{J^2f_2}{J+f_2\omega /f_2},
$$
$$
E_p=\frac{\epsilon _p-J}2-\sqrt{\left( \frac{\epsilon _p+J}2\right)
^2+J^2f_2 }\approx -\frac 32J.
$$
The perturbation theory result is
$$
\Sigma _{pert}\left( p,\omega \right) \approx J^2f_2/\omega ,
$$
$$
E_{pert}=\frac{\epsilon _p}2-\sqrt{\left( \frac{\epsilon _p}2\right)
^2+J^2f_2}\approx -J\sqrt{\frac 34}>\Omega _S.
$$

The reason of the perturbation theory fail is the violation of the sum rule
( \ref{sigsrule2}). We have
$$
\rho _{pert}(p,\omega )=\frac{J^2}N\sum_qC_q\delta (\omega -\epsilon
_{p+q}),
$$
$$
{\rm a}_{1,pert}=\frac 1{{\rm b}_1}\frac{J^2}N\sum_qC_q\int_{-\infty
}^{+\infty }\omega \delta (\omega -\epsilon _{p+q})d\omega =
$$
\begin{equation}
\label{a1pert}=\frac 1{f_2N}\sum_qC_q\epsilon _{p+q}=\frac{C_g}{f_2}\epsilon
_p.
\end{equation}
Comparing (\ref{a1pert}) with the exact value given by Eq. (\ref{matel}) we
see the absence of terms proportional to $J$ that give completely wrong
result for $J\gg t$. It is not difficult to prove that summation of the
infinite series of non crossing diagrams for the self energy does not change
the value of ${\rm a}_{1,pert}$ (\ref{a1pert}). Indeed, the non-crossing
(self-consistent Born) approximation gives
\begin{equation}
\label{noncr}\Sigma _{n-c}(p,\omega )=\frac{J^2}N\sum_qC_qG_a\left(
p+q,\omega -\omega _q\right) ,
\end{equation}
where $\omega _q$ is the energy of spin excitations. We then have
$$
{\rm a}_{1,n-c}=\frac 1{{\rm b}_1}\frac{J^2}N\sum_qC_q\int_{-\infty
}^{+\infty }\omega A_{aa}\left( p+q,\omega -\omega _q\right) d\omega =\frac 1
{f_2N}\sum_qC_q\left( \epsilon _{p+q}+\omega _q\right) ,
$$
and obtain the same result (\ref{a1pert}) because $\omega _q$ is negligible,
at least in the vicinity of $q=Q$.

Moreover, the self-energy (\ref{noncr}) leads to the absence of
quasiparticles. We may write $\Sigma _{n-c}(p,\omega )\approx
J^2f_2G_a\left( p+Q,\omega \right) $, then
$$
G_a(p,\omega )=\left[ \omega -\epsilon _p-J^2f_2G_a(p+Q,\omega )\right]
^{-1}=\left[ \omega -\epsilon _p-\frac{J^2f_2}{\omega -\epsilon
_{p+Q}-J^2f_2G_a(p,\omega )}\right] ^{-1}.
$$
Solution of the quadratic equation gives the expression for Green's function
$$
G_a(p,\omega )=\frac{(\omega -\epsilon _p)(\omega -\epsilon _{p+Q})-\sqrt{
(\omega -\epsilon _p)^2(\omega -\epsilon _{p+Q})^2-4J^2f_2(\omega -\epsilon
_p)(\omega -\epsilon _{p+Q})}}{2J^2f_2(\omega -\epsilon _p)},
$$
which has no poles. Analogous result was obtained previously in Ref.\cite
{perlov}.

\section{Numerical results}

On Figure 1a,b we present the spectral density $A_{aa}(p,\omega +i\eta
),\eta =0.05t$ that corresponds to three different representation of fermion
self-energy: two pole approximation , Eq. (\ref{sig2}), generalized
projection method, Eq. (\ref{sigMax}), and second order perturbation theory,
Eq. (\ref{sigpert}). We took the value $J/t=3$ that is typical for
underdoped cuprates \cite{Pines,Chub,MMP}. For spin-spin correlation
function we took the expression
$$
C_q=\sqrt{\frac{3|C_g|(1-\gamma _q)}{2\alpha _1(1+\gamma _q)}},
$$
which is provided by spherically symmetric theory for Heisenberg model on
square lattice \cite{Shimah,Starykh} ($\gamma _q\equiv (\cos q_x+\cos
q_y)/2,C_g\approx -0.35,\alpha _1\approx 2.35$). We calculate the function $
D_p(\omega )$ by direct summation over $n\times n$ $q-$points in Brillouin
zone (the results for $n=32$ and $n=64$ are almost indistinguishable). From
Fiq.1 we see that the lowest pole position obtained by generalized
projection method satisfy the relation (\ref{bound}), in contrast to that
given by perturbation theory.

Figure 2 shows the spectral function in generalized projection method for
various values of $p$ along the diagonal of Brillouin zone. Quasiparticle
pole exists throughout the whole Brillouin zone. Quasiparticle dispersion $
E_p$ has only approximate symmetry relative to the boundary of
antiferromagnetic Brillouin zone. Strong asymmetry of the spectral weight in
the generalized projection method is the consequence of the sumrule (\ref
{a0A}) - center of gravity $A_{aa}(p,\omega )$ should coincide with ${\rm a}
_0=\epsilon _p$. So, near the point $p=(\pi ,\pi )$ , where the
quasiparticle peak is far from $\epsilon _p$ , its weight is small.

We have shown above that the expression (\ref{sigMax}) gives reasonable
solutions both in weak and strong coupling limits. Thus we may expect it to
be valid in the intermediate regime $J\sim t$. Figure 3 demonstrate the
fermion spectral density for $J=1.5t$ . In this regime the quasiparticle
solution exists only near the two band minima $p=(\pi ,\pi )$ and $p=(0,0)$
. In other points (for which we could say that quasiparticle pole lies
within the band of bare fermions) the interaction mixes solutions with
different $p$. For $p=0.2(\pi ,\pi )$ and $p=0.8(\pi ,\pi )$ we have
resonant solutions near the bottom ($\epsilon _{\min }=-4t$) and the top ($
\epsilon _{\max }=4t$) of bare fermionic band respectively. Near $p=(\pi
/2,\pi /2)$ we have pure incoherent spectrum.

\section{Conclusion}

We have considered spin-fermion model that is often used for description of
strongly correlated systems. We have compared two popular approaches to the
calculation of fermionic Green's function: the non-crossing approximation of
perturbation theory and Mori-Zwantzig projection technique. We have shown
that first one is valid only in the weak coupling regime and the second one
only in the strong coupling regime. The non-crossing approximation gives
wrong position of lowest quasiparticle band. The reason is the rough
violation of the sum rule for the third moment of spectral density. The
summation of the infinite series of non-crossing diagrams for the
self-energy does not alter this result. We have proposed the generalized
version of projection method that treat exactly the quadratic on fermion
fields the kinetic energy term. The resulting expression for the self-energy
coincides with that of perturbation theory in weak coupling limit and
provides right quasiparticle pole position in strong coupling limit. We thus
consider it as a good starting point to the investigation of intermediate
coupling regime that is believed to be relevant for optimally doped cuprate
compounds.

\section{ACKNOWLEDGMENTS}

This work was supported, in part, by the INTAS-RFBR (project No. 95-0591).
No.\ 95-0591), by RSFR (Grant No. 98-02-17187 and 98-02-16730) and by
Russian National program on Superconductivity (Grant No. 93080). R.O.K.
thanks the Institute for High Pressure Physics for hospitality during
accomplishing of part of this work.

\begin{figure}
\caption{The spectral density of the one particle Green's function with the
self-energy, obtained by three different ways: two pole approximation
(\protect\ref{sig2}), generalized
projection method, (\protect\ref{sigMax}), and second order perturbation
theory, (\protect\ref{sigpert})}

\label{f1}

 \end{figure}

\begin{figure}
\caption{The spectral density in generalised projection method as a function
of quasimomentum $p$ in strong coupling regime $J/t=3$.}

\label{f2}

 \end{figure}

\begin{figure}
\caption{The spectral density in intermediate paremeter regime.}

\label{f3}

 \end{figure}

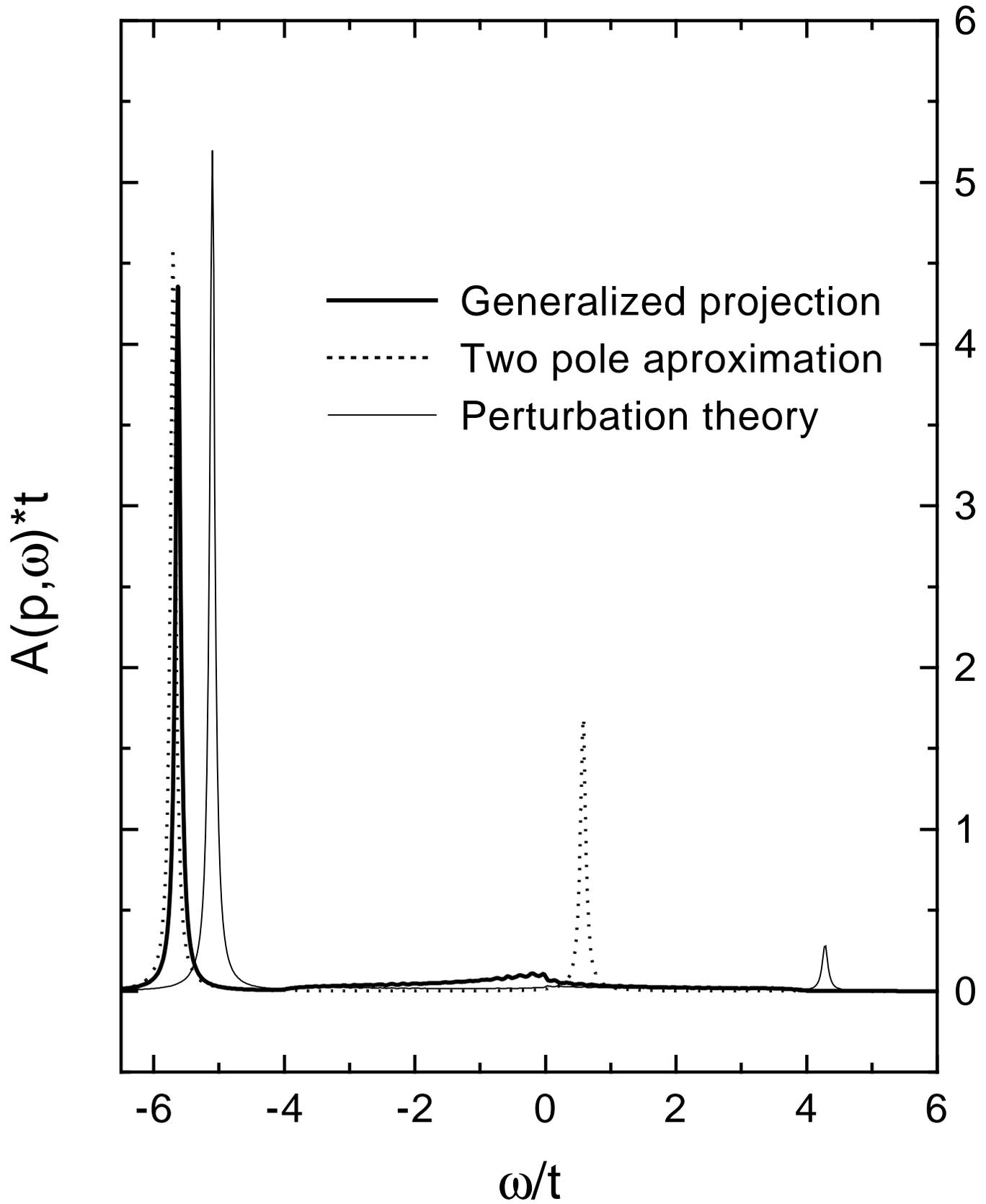
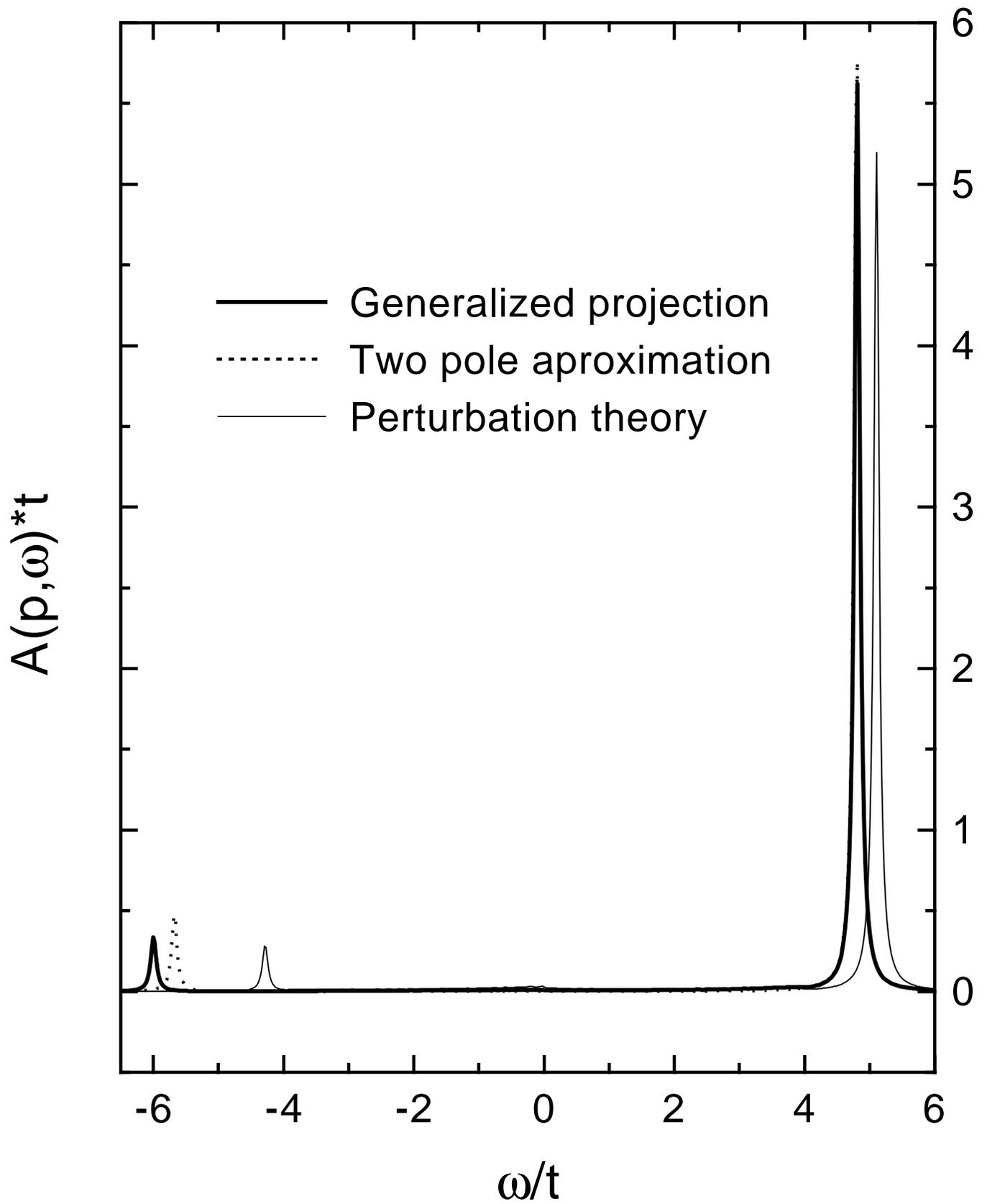
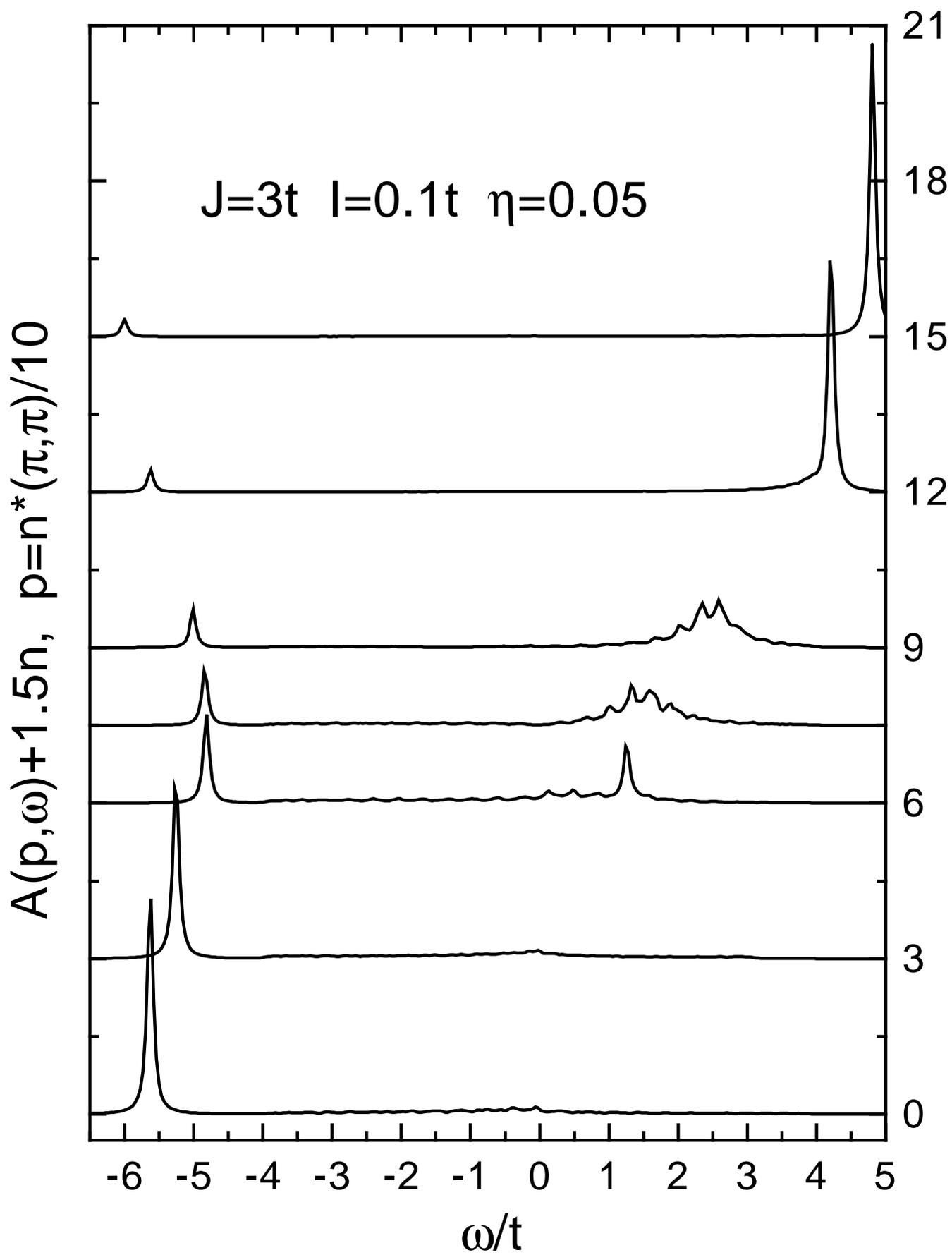
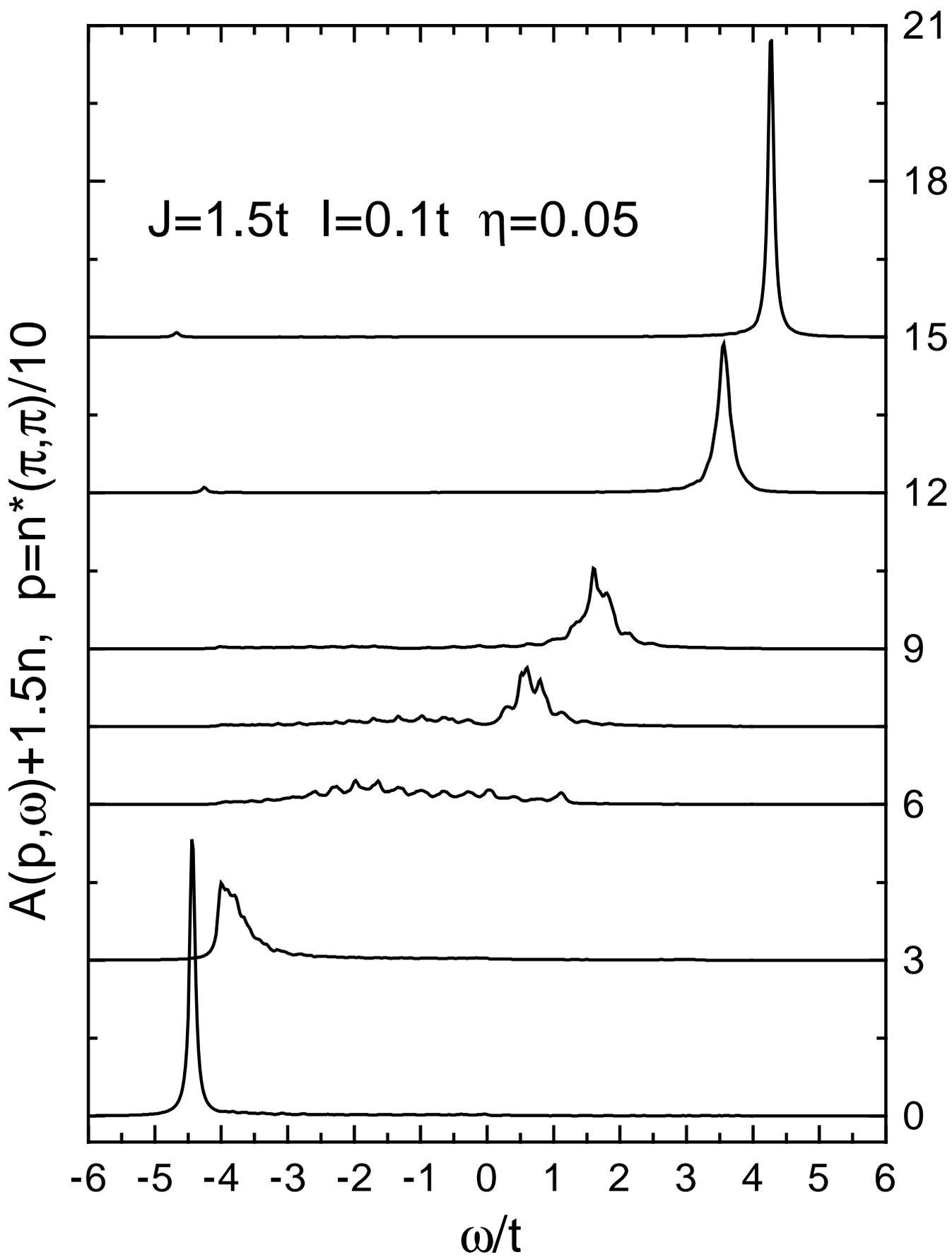
\end{document}